\documentclass[reviewcopy]{elsarticle}

\usepackage[reviewcopy]{adndt}
\usepackage{longtable}
\usepackage{comment}
\usepackage{amsmath}
\usepackage{amssymb}
\usepackage{xcolor}
\biboptions{square,sort&compress}
\bibpunct[]{[}{]}{,}{n}{}{;}
\begin{document}

\begin{frontmatter}

\journal{Atomic Data and Nuclear Data Tables}

\title{\normalfont\textsc{Radiative Transition Properties of Singly Charged Magnesium, Calcium, Strontium and Barium Ions}}

\author{Mandeep Kaur$^a$}
\author{Danish Furekh Dar$^a$}
\author{B. K. Sahoo$^b$}
\author{Bindiya Arora$^a$}

\ead{E-mail: bindiya.phy@gndu.ac.in}
\address{$^{a}$Guru Nanak Dev University, Amritsar, Punjab-143005, India.}
\address{$^{b}$Atomic, Molecular and Optical Physics Division, Physical Research Laboratory, Navrangpura, Ahmedabad-380009, India.}
\cortext[cor1]{Guru Nanak Dev University, Amritsar, Punjab-143005}

\date{today} 

\begin{abstract}  
Accurate values of electric dipole (E1) amplitudes along with their uncertainties for a number of transitions among low-lying states of Mg$^+$, Ca$^+$, Sr$^+$, and 
Ba$^+$ are listed by carrying out calculations using  a relativistic all-order many-body method. By combining  experimental wavelengths 
with these amplitudes, we quote transition probabilities, oscillator strengths and lifetimes of many short-lived excited states of the 
above ions. The uncertainties in these radiative properties are also quoted. We also give electric quadrupole (E2) and magnetic dipole (M1) amplitudes of the metastable states of the Ca$^+$, Sr$^+$, and
Ba$^+$ ions by performing similar calculations. Using these calculated E1, E2 and M1 matrix elements, we have estimated the transition probabilities,
oscillator strengths and lifetimes of a number of allowed and metastable states. These quantities are further compared with the values available
from the other theoretical studies and experimental data in the literature. These data will be immensely useful for the astrophysical observations, 
laboratory analysis and simulations of spectral properties in the above considered alkaline-earth metal ions.
\end{abstract}


\end{frontmatter}

\tableofcontents
\listofDtables
\listofDfigures
\vskip5pc

\section{Introduction}
Spectroscopic properties of singly charged alkaline-earth metal ions are very demanding in many areas of physics. In recent years, these ions
have emerged as the potential candidates to perform high precision experiments of fundamental importance. This is owing to the fact that they
provide long observational time in addition to development of innovative techniques to trap and laser cool them. As a result, precise measurements 
of lifetimes of atomic states~\cite{olms}, light shifts~\cite{sherman}, branching ratios~\cite{kurz2008} and many other fundamental properties
\cite{roos,fortson, sahoo2011} of these ions have been carried out. This has also led to consider these ions for quantum state manipulation
experiments \cite{nigg, wilson}, making atomic clocks~\cite{chou}, studying parity non-conservation effects~\cite{fortson}, probing Lorentz symmetry 
violations \cite{wood, tiecke}, investigating nuclear charge radii \cite{ruiz2016, shi2017} etc.. Frequency standards based on the optical transitions of laser
cooled singly charged Ca$^+$ \cite{huang,champenois} and Sr$^+$ \cite{barwood,fordell,dube} ions are expected to reach systematic fractional uncertainties of 
$10^{-18}$. All these high-precision experiments require accurate estimate of systematics due to the stray electromagnetic radiations in the 
experiments; mainly due to the Stark, Zeeman, quadrupole and black-body radiation shifts. These effects can be estimated with the knowledge of 
electromagnetic transition amplitudes. Another important application that demands accurate values of radiative properties is in astrophysics, where
abundance of different atomic elements can be found out by investigating emission and absorption spectra of galaxies and disks surrounding the stars 
and other stellar objects \cite{welty, mashonkina, persson, hobbs}. 

In recent years, several groups have carried out calculations of radiative properties such as the transition rates, oscillator strengths, lifetimes 
and branching ratios of several states of alkaline-earth-metal ions using a variety of many-body methods both in the non-relativistic and 
relativistic frameworks. Semi-empirical calculations were performed by Theodosiou to determine lifetimes of the (5-7)$S_{1/2}$, (4-6)$P_{3/2,1/2}$, 
(4,5)$D_{5/2,3/2}$ and 4$F_{7/2,5/2}$ states of the alkaline-earth metal ions~\cite{theodosiou1989}. The oscillator strengths due to electric dipole 
(E1) amplitudes of the $S-P$, $P-D$, and $D-F$ transitions in the Mg$^+$, Ca$^+$, Sr$^+$ and Ba$^+$ ions were determined using the scaled 
Thomas-Fermi-Dirac (TFD) wave functions with the spin-orbit interactions by Warner ~\cite{warner1968}. The relativistic pseudo-potential approach was
adopted by Hafner and Schwarz ~\cite{hafner1978} for the evaluation of the E1 transition probabilities of many low-lying transitions of the Ca$^+$ 
ion. Calculations of the oscillator strengths for a number of transitions among the low-lying states of the Mg$^+$ and Ca$^+$ and Sr$^+$ ions were 
carried out by Mitroy and Zhang by diagonalizing the semi-empirical Hamiltonian in a large dimension single electron basis~\cite{mitroy, mitroy2008}.
The multi-configuration Hartree-Fock (MCHF) method was employed by Vaeck et al. to obtain the radiative lifetimes and oscillator strengths 
for the different transitions of low-lying levels in Ca$^+$~\cite{vaeck1992}. A systematic study of atomic properties in Ca$^+$ was conducted by 
Safronova et al. using a relativistic all-order method~\cite{safsaf2011}. Relativistic coupled-cluster (RCC) method was applied to calculate the 
lifetimes, ionization potentials and polarizabilities of the $4S$ and $3D$ states of this ion by Sahoo et al.~\cite{sahoo2009}. There are 
 many other groups who have reported transition probabilities for some of the important transitions of Ca$^+$ at different level of 
approximations ~\cite{augilera, wiese, kurucz, melendez,kramida, guet1991, liaw95, arora2007}. Using the Super Structure code, Bautista 
et al. have evaluated oscillator strengths, transition rates, and electron-impact excitation-rate coefficients for Sr$^+$
\cite{bautista2002}. A relativistic Hartree-Fock (HF) method with core-polarization potential calculations of the oscillator strengths of the lower 
$5S_{1/2}-5P_{3/2,1/2}$ transitions in Sr$^+$ and $6S_{1/2}-6P_{3/2,1/2}$ transitions in Ba$^+$ have been presented by Migdalek and 
Baylis~\cite{migdalek1979}. Lifetime calculations for the $6P_{3/2,1/2}$ states of Ba$^+$ were carried out by Sahoo et al.~\cite{sahoo} 
and Safronova et al.~\cite{safronova2011}.

There have been many measurements of the radiative properties of the singly charged alkaline-earth-metal ions reported by many groups 
around the world. Measurements of the transition probabilities of many spectral lines of Ca$^+$ have been performed by Aguilera et al. 
using the laser-induced breakdown spectroscopy based CSigma graphs~\cite{augilera}. The transition probability for the $4P_{1/2}-4S_{1/2}$ transition of this ion has also been given by Hettrich et al.~\cite{hettrich2015} in 2015 by comparing measurements of
dispersive and absorptive light ion interactions. Also for Sr$^+$, Likforman et al.~\cite{likforman2016} and Zhang et al.~\cite{zhang2016} reported the transition probabilities of the $5P-5S$ and $5P-4D$ transitions in 2016 by using photon-counting sequential method and nested sequences of population inversion method,  respectively. Absolute transition probabilities of the low-lying transitions in 
Ba$^+$ have been measured by Kastberg et al. using optical nutation method~\cite{kastberg}, Gallagher et al.~\cite{gallagher1967} using Hanle-effect technique and Woods et al. ~\cite{woods2010} using resonant
excitation Stark ionization spectroscopy. Recently, Arnold et al.~\cite{arnold2019} and Zhang et al.~\cite{zhang2020} used laser spectroscopy to measure branching fractions for the decays from the $6P_{1/2}$  and $6P_{3/2}$ states, respectively in Ba$^+$. 
Lifetime measurements of the $3P_{3/2}$ state of Mg$^+$ and $4P_{3/2}$ 
state of Ca$^+$ are carried out by Smith and Gallagher using the Hanle-effect technique~\cite{smith1966}. The lifetime of the $3P_{3/2}$ state of Mg$^+$ is reported by Herrmann et al. in 2009 ~\cite{herrmann2009} using frequency metrology on single trapped ions. The Hanle-effect technique was adopted by Gallagher to measure the lifetimes of the $5P_{3/2,1/2}$ states in Sr$^+$, and the $6P_{3/2,1/2}$ states in the 
Ba$^+$ ion~\cite{gallagher1967}. The beam-foil excitation technique was used by Andersen et al. to measure the transition probabilities of
the $5S-4P$, $4D-4P$ and $5D-4P$ states, and the lifetimes of the $4D$ and $5D$ states in Ca$^+$ ion~\cite{andersen1970}. Later, the influence 
of the cascading effects to these measurements was analyzed and revised values of the lifetimes of the above states were given by Emmoth 
et al.~\cite{emmoth1975}. Lifetime for the $4P_{1/2}$ state of Ca$^+$  was reported by Hettrich et al. ~\cite{hettrich2015}  by comparing measurements of dispersive and absorptive light ion interactions. Also,  Meir et al.~\cite{meir2020} reported a recent lifetime measurement  of the $4P_{3/2}$ state using induced light shift and scattering rate on a single trapped Ca$^+$ ion.  A beam pulsed dye laser experiment was conducted by Kuske et al. to measure the lifetimes of the
resonance levels of the $5P_{3/2,1/2}$ states in Sr$^+$ and the $6P_{3/2,1/2}$ states in Ba$^+$ \cite{kuske1978}. Lifetime measurements using the 
laser-induced fluorescence have been carried out by the group of Pinnington to determine the lifetimes of the $4F_{7/2,5/2}$ and $5P_{3/2,1/2}$ 
levels of Sr$^+$, and the $4F_{7/2,5/2}$, $6P_{3/2,1/2}$ and $7P_{3/2,1/2}$ levels of Ba$^+$~\cite{pinnington1995}. The experimental values of the
lifetimes of the $4P_{3/2}$ and $6P_{3/2}$ states of the Ca$^+$ and Ba$^+$ ions, respectively, are reported by Rosner et al.~
\cite{rosner1997}. Again, the branching ratios of the decay probabilities from the $6P_{3/2}$ state of Ba$^+$ have been measured by Kurz 
et al.~\cite{kurz2008}. The most recent lifetime measurements of the $6P_{3/2,1/2}$ states in Ba$^+$ are given by Arnold et al.~\cite{arnold2019} and Zhang et al. ~\cite{zhang2020}. 

The aforementioned literature data mainly focus on the transitions among the low-lying states, while information on the radiative properties of the 
higher excited states are scarce. In this Data Table, we provide accurate values of the E1 amplitudes for a large number of transitions in the 
Mg$^+$, Ca$^+$, Sr$^+$ and Ba$^+$ alkaline-earth-metal ions by employing a relativistic all-order method. Particularly, we present the E1 matrix 
elements of the $q$P$_{3/2,1/2}-q'$S$_{1/2}$ transitions with $q=3-5$ and $q'=3-6$, the $m$D$_{5/2,3/2}- m'$P$_{3/2,1/2}$ transitions 
with $m=3-6$ and $m'=3-5$, and the $n$F$_{7/2,5/2}-n'$D$_{5/2,3/2}$ transitions with $n= 4,5$ and $n'=3-6$ in Mg$^+$; the 
$q$P$_{3/2,1/2}-q'$S$_{1/2}$ transitions with $q=4-6$ and $q'=4-7$, the $m$D$_{5/2,3/2}- m'$P$_{3/2,1/2}$ transitions with $m=3-6$ and $m'=4-6$, and the $n$F$_{7/2,5/2}-n'$D$_{5/2,3/2}$ transitions with $n= 4,5$ and $n'=3-6$ 
in Ca$^+$; the $q$P$_{3/2,1/2}-q'$S$_{1/2}$ transitions with $q=5-8$ and $q'=5-8$, the $m$D$_{5/2,3/2}- m'$P$_{3/2,1/2}$ transitions with $m=4-7$ and 
$m'=5-8$, and the $n$F$_{7/2,5/2}-n'$D$_{5/2,3/2}$ transitions with $n= 4,5$ and $n'=4-7$ in Sr$^+$; and the $q$P$_{1/2,3/2}-q'$S$_{1/2}$ transitions with 
$q=6-8$ and $q'=6-9$, and the $m$D$_{5/2,3/2}- m'$P$_{3/2,1/2}$ transitions with $m=5-8$ and $m'=6-8$ in Ba$^+$. Along with, we have calculated 
the electric quadrupole (E2) and the magnetic dipole (M1) amplitudes between the transitions involving the ground and the metastable states of the 
Ca$^+$, Sr$^+$ and Ba$^+$ ions. Combining all the above E1, E2 and M1 matrix elements with the experimental wavelengths (by deriving from the 
experimental energies), we have determined the oscillator strengths, the transition probabilities and the lifetimes of the metastable states and
many excited states of the above ions. We list these values along with other available experimental and theoretical results in different tables.

In Sec. \ref{sec2}, we present theoretical formulae for evaluating the transition probabilities, the oscillator strengths and the lifetimes of 
atomic states. We also briefly discuss about the relativistic all-order method employed by us for calculating the transition matrix elements in this section. {In Sec. \ref{matel}, we discuss the approach for giving the recommended transition matrix elements and the uncertainties associated with them.} 
All the obtained data, results from the earlier works and discussions are presented in Sec. \ref{results}. This follows by conclusion in 
Sec. \ref{cons}. All the results are given in atomic units (a.u.) unless otherwise stated explicitly.

\section{Theory and Method of Calculation} \label{sec2}

We give below the general formulae for the transition probabilities and oscillator strengths for the E1, E2, and M1 decay channels, and 
the lifetimes of atomic states. Then, we mention about the procedures adopted for calculating atomic wave functions and the transition 
matrix elements.

\subsection{Transition probabilities, oscillator strengths and lifetimes}

The E1 transition probability ($A^{E1}_{vk}$) in inverse second (s$^{-1}$) from an upper level $|\Psi_v\rangle$ with angular momentum $J_v$ to a 
lower level $|\Psi_k \rangle$ with angular momentum $J_k$ in terms of the fundamental constants is given by \cite{kelleher}
\begin{equation}
A^{E1}_{vk}= \frac{2}{3} \alpha c \pi \sigma \times \left (\frac{\alpha \sigma}{R_{\infty}} \right)^{2} \times \frac{S^{E1}}{g_v}, 
\end{equation}
where $R_{\infty} = \frac{\alpha^{2}m_{e} c}{2h}$ is the Rydberg constant, $\alpha  = \frac{e^{2}}{4 \pi \epsilon_{0} \hbar c}$ is the fine structure 
constant, $c$ is the speed of light and {$\sigma=E_v-E_k$ is the energy difference between the upper ($E_v$) and lower ($E_k$) levels of the 
transition}, and $S^{E1}=|\langle J_v||{\bf D}|| J_k \rangle|^2$ is the line strength with ${\bf D}=\sum_j {\bf d}_j 
= - e \sum_j {\bf r}_j$ being the electric dipole (E1) operator with position of $j^{th}$ electron ${\bf r}_j$. {By substituting the values of fundamental constants and wavelength ($\lambda$) of the transition in \AA ,}  $A^{E1}_{vk}$ is conveniently determined using 
the following formula \cite{sobelman1979atomic}  
\begin{equation}
A^{E1}_{vk}= \frac{2.02613 \times 10^{18}}{g_v \lambda^3} \times S^{E1} , \label{e1coeff}
\end{equation}
where, the degeneracy factor $g_v=2J_v+1$ for the corresponding state and line strength 
$S^{E1}$ is in atomic units (a.u.).

Similarly, the E2 transition probability ($A^{E2}_{vk}$) between the $|\Psi_v\rangle$ and $|\Psi_k \rangle$ states is given by 
\begin{equation}
A^{E2}_{vk}=\frac{1}{120} \alpha c \pi \sigma \times \left (\frac{\alpha \sigma}{R_{\infty}} \right)^{4} \times \frac{S^{E2}}{g_v} ,
\end{equation}
with the E2 line strength $S^{E2}=|\langle J_v||{\bf Q}|| J_k \rangle|^2$ for the E2 operator ${\bf Q}=\sum_j {\bf q}_j=
-\frac{e}{2} \sum_j (3z^2_j-r_j^2)$. Using $\lambda$ in \AA ~and $S^{E2}$ in a.u, it can be estimated 
using the formula
\begin{equation}
A^{E2}_{vk} =\frac{1.1199 \times 10^{18}}{g_v\lambda^5} \times S^{E2}.  \label{e2coeff}
\end{equation}

Also, the M1 transition probability ($A^{M1}_{vk}$) is given by
\begin{equation}
A^{M1}_{vk}=\frac{1}{6} \alpha^{3} c \pi \sigma \times \left (\frac{\alpha \sigma}{R_{\infty}} \right)^{2} \times \frac{S^{M1}}{g_v}
\end{equation}
with the M1 line strength $S^{M1}=|\langle J_v||{\bf {\cal M}}|| J_k \rangle|^2$ for the M1 operator 
${\cal M}
=\sum_j {\bf \varsigma}_j$
=$\sum_j ({\bf l}_j+2{\bf s}_j)\mu_{B}$. Here, ${\bf l}_j$ and ${\bf s}_j$ are the orbital and spin angular momentum operators for the $jth$ electron, respectively,  and  $\mu_{B}$ is the Bohr magneton. By converting 
$\lambda$ in \AA ~and $S^{E2}$ in a.u, we can determine $A^{M1}_{vk}$ as
\begin{equation}
A^{M1}_{vk}=\frac{2.69735 \times 10^{13}}{g_v\lambda^3} \times S^{M1} . \label{m1coeff}
\end{equation}

Using the above transition probabilities, the absorption oscillator strengths $f_{kv}^O$ for the transition operator $O$ representing the 
corresponding E1, E2 and M1 operators are calculated as~\cite{sobelman1979atomic, kelleher} 
\begin{equation}
f_{kv}^{O}= \left(\frac{R_{\infty}}{2c \alpha^{3} \pi} \right)\frac{g_v}{g_k} \times \frac{A^{O}_{vk}}{\sigma^{2}} = 1.4992 \times10^{-16}\times 
\frac{g_v}{g_k} A^{O}_{vk} \lambda^{2} . \label{eqsf}
\end{equation}
This follows that
\begin{eqnarray}
f^{E1}_{kv} &=& \frac{1}{3\alpha}\left(\frac{\alpha \sigma}{R_{\infty}}\right) \times \frac{S^{E1}}{g_k} = \frac{303.756}{g_k \lambda} 
\times S^{E1}, \label{eq-os} \\
f^{E2}_{kv}&=&\frac{1}{240\alpha}\left(\frac{\alpha \sigma}{R_{\infty}}\right)^{3} \times\frac{S^{E2}}{g_k} 
= \frac{167.90}{g_k \lambda^3} \times S^{E2} \label{eq-osq}
\end{eqnarray}
and
\begin{eqnarray}
f^{M1}_{kv}= \frac{\alpha}{12}\left(\frac{\alpha \sigma}{R_{\infty}}\right) \times \frac{S^{M1}}{g_k}=\frac{0.00404386}{g_k\lambda} \times S^{M1} . \label{eq-osm}
\end{eqnarray}
It can be noted that the emission oscillator strength for a transition can be deduced from its absorption oscillator strength due to the 
corresponding transition operator $O$ by using the relation 
\begin{equation}
f_{vk}^O=-\frac{g_k}{g_v}f_{kv}^O .
\end{equation}
For the evaluations of the above quantities, we have used the values of the fundamental constants as $\alpha  = 7.297352\times10^{-3}$, 
$c=29979245800$ cm s$^{-1}$ and $R_{\infty} =1.0973731\times 10^5$ cm$^{-1}$ from Ref.~\cite{mohr}. 

After determining transition probabilities due to all possible decay channels from a state $|\Psi_v\rangle$, its lifetime ($\tau_v$) is estimated by
\begin{equation}
\tau_v=\frac{1}{\sum_{O,k} A^O_{vk}}
\end{equation}
in s, where sum over $O$ means all possible decay channels and sum over $k$ represents all possible lower states than the states for which lifetime
is estimated. As it can be observed from Eqs. (\ref{e1coeff}) to (\ref{m1coeff}), transition probability due to E1 channel contributes predominantly 
as compared to the other two channels (several orders magnitudes higher). Therefore, we neglect contributions due to E2 and M1 transition probabilities in the 
estimations of lifetimes of the short-lived excited states. However, the lifetimes of the metastable states are evaluated from the combined transition 
probabilities of all possible E2 and M1 channels. 

\subsection{Relativistic all-order method}

For accurate evaluation of the transition matrix elements, we determine atomic wave functions by using a relativistic  all-order method. We refer 
to Refs. \cite{Blundell,theory, PhysRevA.91.042507,PhysRevA.92.052511} for the detailed description of this method, which is also described briefly 
here. In this approach, wave function of an atomic state having a closed-shell configuration and a valence orbital is expressed by considering 
singles and doubles excitations (SD method) as
\begin{eqnarray}
|\Psi_v \rangle_{\rm SD} &=& \left[1+ \sum_{ma}\rho_{ma} a^\dag_m a_a+ \frac{1}{2} \sum_{mnab} \rho_{mnab}a_m^\dag a_n^\dag a_b a_a + 
\sum_{m \ne v} \rho_{mv} a^\dag_m a_v + \sum_{mna}\rho_{mnva} a_m^\dag a_n^\dag a_a a_v\right] |\Phi_v\rangle,
 \label{expansion}
\end{eqnarray}
where $|\Phi_v\rangle$ is the mean-field wave function constructed as $|\Phi_v\rangle=a_v^{\dag}|0_c\rangle$ with $|0_c\rangle$ representing the 
Dirac-Hartree-Fock (DHF) wave function, the second quantization $a^\dag_i$ and $a_i$ operators represent as the creation and annihilation operators
with the indices $m,n, \cdots$ and  $a,b \cdots$ referring to the unoccupied and occupied orbitals, respectively, while the index $v$ designates for 
the valence orbital. In the above expression, $\rho_{ma}$ and $\rho_{mv}$ correspond to the single hole-particle and valence-particle excitation 
coefficients, respectively, and $\rho_{mnab}$ and $\rho_{mnva}$ are the double hole-particle and valence-hole--particle excitation coefficients,
respectively. These excitation coefficients are obtained by solving the Dirac equation including the electron-nucleus and electron-electron 
Coulomb interactions self-consistently in an iterative procedure. The resulting expansion coefficients are further used to calculate the matrix elements. 
Contributions from the next leading order effects are  evaluated using partial triple excitations in the SD method (SDpT method) by defining
\begin{equation}
|\Psi_v\rangle_{\rm SDpT}\approx |\Psi_v\rangle_{\rm SD}+\left[\frac{1}{6}\sum_{mnrab}\rho_{mnrvab}a_m^{\dag}a_n^{\dag}a_r^{\dag}a_ba_aa_v+
\frac{1}{18}\sum_{mnrabc}\rho_{mnrabc}a_m^{\dag}a_n^{\dag}a_r^{\dag}a_ca_ba_a\right] |\Phi_v\rangle,
\label{expansion1}
\end{equation}
where the triple excitation $\rho_{mnrvab}$ and $\rho_{mnrabc}$ coefficients are determined in a perturbative procedure. To obtain the DHF 
wave function, we use 70 B-splines of order $k = 11$ for each angular momentum. The basis set orbitals are defined on a non-linear 
grid and are constrained to a large spherical cavity of a radius $R = 220$ a.u.. This choice of spline and cavity radius ensures accommodation of 
all valence orbitals considered in this work.

\subsection{Evaluation of matrix elements}

The matrix element of a one-body operator $O$ between the states $|\Psi_v\rangle$ and $|\Psi_w\rangle$ is evaluated as
\begin{equation}
 O_{vw} = \frac{\langle \Psi_v|O|\Psi_w\rangle}{\sqrt{\langle\Psi_v|\Psi_v\rangle\langle\Psi_w|\Psi_w\rangle}},
 \label{wavef1}
\end{equation}
where $O$ corresponds to either of the E1, E2 or M1 operators. After substituting SD and SDpT form of wave functions given by Eqs. (\ref{expansion})
and (\ref{expansion1}), respectively, in the above expression, we obtain results as explicit sum of core, core-valence and valence correlation 
contributions. The numerator does not have any contribution from core correlation as the states involved in these matrix element calculations 
have different valence orbital. It can be divided into the DHF term and 20 other terms containing electron correlation effects that are either linear 
or quadratic functions in excitation coefficients~\cite{CC}. Among them, only two terms give dominant contributions to nearly all the transitions
\begin{equation}
O_{vw}^{(a)}=\sum_{ma}(o_{am}\widetilde{\rho}_{vmwa}+o_{ma} \widetilde{\rho}^*_{wmva})
\end{equation}
and
\begin{equation}
O_{vw}^{(c)}=\sum_m(o_{vm}\rho_{mw}+o_{mw}\rho^*_{mv}),
\end{equation}
where $\tilde{\rho}_{vmwa}=\rho_{vmwa}-\rho_{mvaw}$ and $*$ means complex conjugate term. In the many-body theory terminology, $O_{vw}^{(a)}$ and $O_{vw}^{(c)}$ account for the electron  
core-polarization and Br\"uckner pair-correlation effects respectively. In the SD approximation, $O_{vw}^{(a)}$ includes core-polarization effects to 
all-orders like the random phase approximation. However, $O_{vw}^{(c)}$ is complete only up to third-order but it misses out some of the important 
contributions corresponding to the fourth-order relativistic many-body perturbation theory (RMBPT). Using the wave functions from the SDpT method, 
these missing contributions are included perturbatively. In the end, the other omitted correlation contributions are estimated by scaling the 
calculated wave functions by analyzing the ratio of the experimental to theoretical correlation energies. We present results as $O^{\rm {DHF}}$, 
$O^{\rm {SD}}$ and $O^{\rm{SDpT}}$ by using wave functions from the DHF, SD, and SDpT methods, respectively. When the scaled wave functions from the 
SD and SDpT methods are used for the matrix element evaluations, the corresponding results are quoted as $O^{\rm{SD}}_{\rm{sc}}$ and 
$O^{\rm{SDpT}}_{\rm{sc}}$, respectively. We estimate the recommended values for matrix elements by comparing the ratio $R=O_{vw}^{(c)}/O_{vw}^{(a)}$.
For $R$ \textgreater 1, results from the SD scaled wave functions are considered as the final values; otherwise results from the SD calculations 
are given as the final values. More details on this procedure can be found in Refs. \cite{CC, Blundell}. These final values are quoted as 
$O^{\rm {Final}}$ in different tables.

\section{E1, E2 and M1 matrix elements}~\label{matel}

The E1 matrix elements calculated using the DHF, SD, SDsc, SDpT and SDpTsc methods for a number of $P-S$, $D-P$ and $F-D$ transitions of the Mg$^+$, 
Ca$^+$, Sr$^+$ and Ba$^+$ alkaline-earth metal ions are listed in Tables \ref{matrixmg}, \ref{matrixca}, \ref{matrixsr} and \ref{matrixba}, 
respectively. Since the present work focuses mostly on the allowed transitions, reliability in the calculations of E1 matrix elements are verified 
by performing calculations by using both length and velocity gauge expressions. The single particle wave function for a orbital $v$ (denoted by 
$|j_vm_{j_v}\rangle$ with orbital angular momentum $j$ and its azimuthal quantum number $m_{j_v}$) in the relativistic theory has the form
\begin{eqnarray}
 |j_v m_{j_v} \rangle = \frac{1}{r} \left ( \begin{matrix} iG_v(r) |\chi_{\kappa_v,m_{j_v}}(\theta, \phi) \rangle \\  F_v(r) |\chi_{-\kappa_v,m_{j_v}}(\theta, \phi) \rangle \end{matrix} \right ) ,
\end{eqnarray}
where $G_v(r)$ and $F_v(r)$ are the large and small component of the Dirac wave function, respectively. Here, $|\chi_{\kappa_v,m_{j_v}}(\theta, \phi)\rangle$  
with relativistic angular momentum $\kappa_v$ corresponds to angular factor that takes into account the spin-orbit coupling and is responsible of 
ensuring selection rules in the calculations. In terms of the single particle orbital wave functions, the reduced matrix element of the E1 
operator between $|j_vm_{j_v}\rangle$ and $|j_wm_{j_w}\rangle$ is given by
\begin{eqnarray}
\langle j_v||d||j_w\rangle&=&\frac{3}{k}\langle\kappa_v||C_1||\kappa_w\rangle\int_0^{\infty}dr\left\{j_1(kr)[G_v(r)G_w(r)+F_v(r)F_w(r)\right.\nonumber\\
&&+\left.j_2(kr)\left[\frac{\kappa_v-\kappa_w}{2}[G_v(r)F_w(r)+F_v(r)G_w(r)]+[G_v(r)F_w(r)-F_v(r)G_w(r)]\right]\right\}
\end{eqnarray}
in the length gauge and
\begin{eqnarray}
\langle j_v||d||j_w\rangle&=&\frac{3}{k}\langle\kappa_v||C_1||\kappa_w\rangle\int_0^{\infty}dr\left\{-\frac{\kappa_v-\kappa_w}{2}\left[\frac{dj_1(kr)}{d kr}+\frac{j_1(kr)}{kr}\right]\right.\nonumber\\
&& \times \left.[G_v(r)F_{w}(r)+G_v(r)F_w(r)]+\frac{j_1(kr)}{kr}[G_v(r)F_w(r)-F_v(r)G_w(r)]\right\}
\end{eqnarray}
in the velocity gauge. In the above expression, $k=\omega \alpha$ with the emitted photon energy $\omega$ of the transition, $C_{n}$ is the 
normalized spherical harmonic of rank $n$ and $j_l(x)$ is a spherical Bessel function of order $l$, which is given by
\begin{eqnarray}
 j_l(x) = \frac{x^l}{1 \cdot 3 \cdot 5 \cdots (2l+1)} .
\end{eqnarray}
About 84 number of E1 transitions each for Mg$^+$ and Ca$^+$, 104 number of E1 transitions for Sr$^+$ and 60 number of E1 transitions for Ba$^+$ are 
calculated in this work. We have compared results from both the length and velocity gauge expressions using the SD and SDpT methods in the above 
tables. A very good agreement is seen among both set of results for most of the cases except in a very few transitions involving the high-lying excited
states. This can be understood as it is well known that convergences in the velocity gauge results are usually achieved after including a large 
configuration space in the calculations. Thus, it would be imperative to take into account contributions from triple excitations to get better match 
between both gauge results, but  it is beyond the scope of present work. We also list our final recommended values for these E1 matrix elements along
with their estimated uncertainties in the same tables. We have established these uncertainties based on the ratio $R$~\cite{rb2011}. If $R$ lies in
range 0.5 \textless $R$ \textless 1.5, then uncertainty is evaluated as the maximum difference between the final value of matrix element and the other
three all-order values. However, if 1.5 \textless $R$ \textless 3, then final uncertainty is estimated as max(SDsc - SD, SDsc - SDpT, SDsc - SDpTsc). 
Also, the final uncertainty is calculated as max(SDsc - SDpT, SDsc - SDpTsc) if $R$ \textgreater 3. For a few transitions large differences between the results from the length and velocity gauge expressions are observed. It can be noticed that in such cases large uncertainties have been quoted. This ensures that our approach of estimating uncertainties to the E1 matrix elements are reliable.

We have also evaluated matrix elements only for a few forbidden transitions involving the ground and metastable states of the considered ions. For 
the E2 transitions, the reduced matrix element is evaluated using the length gauge expression given by
\begin{eqnarray}
\langle j_v||q||j_w\rangle &=&\frac{15}{k^2}\langle\kappa_v||C_2||\kappa_w\rangle\int_0^{\infty}dr\left\{j_2(kr)[G_v(r)G_w(r)+F_v(r)F_w(r)]\right.\nonumber\\
&+&\left.j_3(kr)\left[\frac{\kappa_v-\kappa_w}{3}[G_v(r)F_w(r)+F_v(r)G_w(r)]+[G_v(r)F_w(r)-F_v(r)G_w(r)]\right]\right\}. 
\end{eqnarray}
 Similarly, the M1 reduced matrix element is evaluated by using the expression
\begin{equation}
\langle j_v||\varsigma||j_w\rangle=\frac{6}{\alpha k}\langle-\kappa_v||C_1||\kappa_w\rangle\int_0^{\infty}dr\frac{(\kappa_v+\kappa_w)}{2}j_1(kr)[G_v(r)F_w(r)+F_v(r)G_w(r)].
\end{equation}
 In Table~\ref{matrixe2m1}, we list the calculated E2 and M1 matrix elements for the $nD_{5/2,3/2}-(n+1)S_{1/2}$ and $nD_{5/2}-nD_{3/2}$ forbidden  
transitions, where $n=3, 4$ and 5 for Ca$^+$, Sr$^+$ and Ba$^+$, respectively. For the E2 and M1 transitions, we present matrix elements only from the DHF and SD methods and the uncertainties are estimated in the similar procedure as for the E1 transitions.  Since we found the uncertainties are extremely small (beyond the significant digits) for the M1 transitions, we have not quoted them explicitly.

\section{Data analysis and Discussion} \label{results}

We have determined the transition probabilities, oscillator strengths and lifetimes of different states using the transition matrix elements discussed 
above. We have neglected contributions from the forbidden transition probabilities in the estimation of lifetimes of the allowed transitions. All 
these results are discussed below. We have also compared our results with the available experimental and other theoretical results that are reported 
in the literature.

\subsection{E1 transition related properties}

With the aid of recommended set of our E1 matrix elements and experimental energies, we determine the transition wavelength $\lambda$, line strengths
$S_{vk}$, transition probabilities $A_{vk}$ and oscillator strengths $f_{kv}$ for various allowed transitions in the considered alkaline earth-metal ions. 
The results of our calculations are summarized in Tables  \ref{parametersmg}, \ref{parametersca}, \ref{parameterssr} and \ref{parametersba} for Mg$^+$,
Ca$^+$, Sr$^+$ and Ba$^+$ ions, respectively. To further show the accuracy of the present calculations, we compare our $A_{vk}$ and $f_{kv}$
results with National Institute of Standards and Technology Atomic Spectra Database (NIST ASD)~\cite{kramida}. In general, our results show very good
agreement with the results given in the NIST ASD for Mg$^+$ except for $6D-(4,5)F$, $5D-4F$ and $6D-3P$ transitions. 
However, we notice that matrix elements from both the length and velocity gauge expressions are in good agreement with each other, which emphasizes the reliability of our results.
From Table \ref{parametersca}, 
we find that our data agrees with the values available in the NIST ASD for all levels in Ca$^+$ with exception of the $4P_{3/2}-3D_{5/2}$, $5P_{3/2,1/2}-3D_{3/2}$  and 
$6P_{3/2,1/2}-3D_{3/2,5/2}$ transitions. We also notice from Table~\ref{parameterssr} that our results for Sr$^+$ ion are generally in good agreement with 
the results available in Ref.~\cite{kramida} except for the $6P_{3/2}-5S_{1/2}$ transition. A similar disagreement between our numbers from 
Table~\ref{parametersba} and those given in the NIST ASD are observed for the $(6,7,8)P-5D$, $9S-6P$, $8D-6P$ and $7P-6S$ transitions in Ba$^+$.  However, as discussed next in this section, our results
for  the  $6P-5D$ transitions in Ba$^+$ are in very good agreement with other theoretical and experimental literature available.To our knowledge, no data can be found in literature to verify accuracy of results for the other transitions. So, we suggest further experimental and theoretical analysis for the transition probabilities of these transitions in Ba$^+$. Also, comparisons are made for our 
estimated $A_{vk}$ coefficient with some existing experimental data and previous calculations for Mg$^+$, Ca$^+$, Sr$^+$ and Ba$^+$ in 
Tables~\ref{comparmg}, \ref{comparca},~\ref{comparsr} and ~\ref{comparba}, respectively. 

We compare the transition probabilities of Mg$^+$ with available theoretical HF calculations by Usta~\cite{usta}, MCHF calculations by 
Fischer et al.~\cite{fischer}, semi-empirical calculations by Theodosiou et al. \cite{theodosiou1999} and empirical calculations by Sofia et al.~\cite{sofia} in Table~\ref{comparmg}. 
We have found that our estimated values are in perfect accord with these previously reported values. Also, the transition probabilities calculated by Majumder et al.~\cite{majumder2002} by employing RCC theory are in excellent accord with our calculated values except for $5D_{3/2}-5P_{3/2,1/2}$ transitions. Transition probabilities are not explicitly given in this paper. So, these values are obtained from the oscillator strengths values for these transitions given therein.  Along with the theoretical calculations, a very good agreement is found with beam-laser measurements of Ansbacher et al.~\cite{ansbacher1989} and measurement by Herrmann et al.~\cite{herrmann2009} using frequency metrology on single trapped ions for the $3P_{3/2,1/2}-3S_{1/2}$ transitions. Also, the transition probabilities of the $4P_{3/2,1/2}-3S_{1/2}$ transitions agree well with the measurements made by Fitzpatrick~\cite{fitz}. 

In Table~~\ref{comparca}, the results for $A_{vk}$ of Ca$^+$ are compared with the previous calculations from the relativistic all-order
method~\cite{safsaf2011}, RCC method~\cite{sahoo2009}, TFD method ~\cite{wiese,melendez}, RMBPT~\cite{guet1991} and Brueckner approximation
\cite{liaw95}. Our results match well with the results of other relativistic all-order work  by Safronova et al.~\cite{safsaf2011}. We also do not 
notice any significant discrepancies between our results and other calculations. The observed small differences among them may be arising due to 
different approximations made in the employed many-body methods in various works. It can be seen in the above table that for the $5S-4P$, $4D_{3/2}-4P_{1/2}$ and $4D_{5/2}-4P_{3/2}$ transitions, the experimental results by 
Andersen et al.~\cite{andersen1970}, who have used beam-foil excitation method, and Augilera et al.~\cite{augilera} who have used laser-induced
breakdown spectroscopy based on CSigma graphs agree very well with our calculations, except for the results for 
$5D_{5/2}-4P_{3/2}$ transition. But, we notice from the table that the results for this transition from the available two experimental measurements give 
substantially different values. So, it is difficult to press upon the validity of the $A_{vk}$ values for this transition using these experimental results. 
Some disagreements between our results and the experimental results reported by Augilera~\cite{augilera} for the $4D_{3/2}-4P_{3/2}$, $6S_{1/2}-4P_{3/2}$ 
and $5D_{5/2,3/2}-4P_{3/2,1/2}$ transitions are seen. We have been unable to compare our predicted results for these transition with any other 
experimental results because of their unavailability in literature. However, we noticed a reasonable agreement of our numbers with other theoretical 
values for these states. Our estimated value for the $4P_{1/2}-4S_{1/2}$ transition matches well with the recent experimental result by Hettrich et al.~\cite{hettrich2015}. 

The transition probabilities for Sr$^+$ are compared in Table \ref{comparsr} with RMBPT calculations by Guet \textit{et.al.}~\cite{guet1991}. Our 
estimated values are in perfect agreement with these values. Whereas, in the same table, a disagreement can be noticed between our calculations and the 
MCHF calculations by Brage \textit{et.al.}~\cite{brage1998} as well as with the Dirac Fock calculations by Ziltis~\cite{ziltis}. Here, an important point 
to be noticed is that calculations by \cite{brage1998} and \cite{ziltis} also present mutual conflict with each other. Moreover, our numbers for all the transitions are also in very good accord with measurements made by Gallagher~\cite{gallagher1967} using Hanle effect with optical excitations method except for the 
$5P_{3/2}-4D_{3/2}$ transition. Our theoretical values for  the $5P-4D$ and $5P-5S$ transitions are in good accordance with the latest measurements reported by Likforman et al.~\cite{likforman2016} and Zhang et al.~\cite{zhang2016}. Good agreement of our results with all these experimental values demonstrates the accuracy of our calculations. 

In Table~\ref{comparba}, transition probabilities for the $6P-5D$ and $6P-6S$ transitions of Ba$^+$ from our calculations 
exhibit good agreement with the theoretical all-order calculations by Safronova et al.~\cite{safronova2011} and RCC calculations by Gopakumar et 
al.~\cite{gopakumar}. Our values for $6P-5D$ transitions within error bars are close to those of Gopakumar et al.~\cite{gopakumar}. We find that our results agree well with the experimental values given by Gallagher~\cite{gallagher1967} and 
Kastberg~\cite{kastberg} within the reported uncertainties for all the transitions with an exception for the $6P_{3/2}-6S_{1/2}$ transition in 
Ba$^+$. In this case, the theoretical value falls just outside the experimental uncertainty of Kastberg \cite{kastberg}. Moreover, a very important point to be noted is that our theoretical calculations for the $6P-5D$ and $6P-6S$ transitions commensurate with the experimental values within uncertainty limits reported by Woods et al.~\cite{woods2010}, Arnold et al.~\cite{arnold2019} and Zhang et al.~\cite{zhang2020}. This indicates that accuracy of our results for other transitions are of similar order. 

\subsection{E2 transition related properties}
We combine the experimental energies and our final recommended values of the E2 matrix elements to calculate the line strengths $S_{vk}$, transition
probabilities $A_{vk}$ and oscillator strengths $f_{kv}$ for the $nD-(n+1)S$ and $nD_{5/2}-nD_{3/2}$ transitions of Ca$^+$($n=3$), Sr$^+$($n=4$) and 
Ba$^+$($n=5$), and present them in Table~\ref{parameterse2}. In this table, we compare our data with the values mentioned in the NIST ASD 
that are primarily available for the $3D-4S$ and $4D-5S$ transitions of Ca$^+$ and Sr$^+$ ion, respectively. We find that our results for Sr$^+$ are 
in excellent agreement with values provided in Ref.~\cite{kramida}, whereas  significant differences are found for the results in Ca$^+$. But, a very important point to be noticed here is that our estimated quadrupole transition rate for the 
$3D_{5/2}-4S_{1/2}$ transition in Ca$^+$ is in good agreement with the recent experimental results by Shao et al.~\cite{shao2017} within uncertainty limits. This provides strong evidence towards the accuracy of our results.  Also, our estimated values are in very good accord with recent semi-empirical core potential calculations by Filippin et al.~\cite{filippin} for Ca$^+$, Sr$^+$ and Ba$^+$, RCC calculations by Guan et al.~\cite{guan2015} for Ca$^+$ and pseudo relativistic Hartree-Fock method calculations by Gurell et al.~\cite{gurell2007} for Ba$^+$. 

\subsection{M1 transition related properties}

The results for the line strengths $S_{vk}$, transition probabilities $A_{vk}$ and oscillator strengths $f_{kv}$ for several levels of interest 
due to M1 transitions of the metastable states in Ca$^+$, Sr$^+$ and Ba$^+$ are summarized in Table~\ref{parametersm1}. The $A_{vk}$ and $f_{kv}$ 
values for the $3D_{5/2}-3D_{3/2}$ transition of Ca$^+$ is in excellent agreement with the NIST ASD values. Our estimated values are mostly in good agreement with the available data from Filippin et al.~\cite{filippin}, Guan et al.~\cite{guan2015} and Gurell et al.~\cite{gurell2007}. A disagreement is found for the $5D_{3/2}-6S_{1/2}$ transition between our value and values provided by Filippin et al.~\cite{filippin} and Gurell et al.~\cite{gurell2007}.  We also notice a mutual conflict in these values by Filippin et al.~\cite{filippin} and Gurell et al.~\cite{gurell2007}. Hence, the accuracy for the value of this transition can not be emphasized based on this available data. 
\subsection{Lifetimes of atomic states}

\textit{Mg$^+$ ion}: The lifetimes of the (3-5)$P_{3/2,1/2}$, (4-6)$S_{1/2}$, (3-5)$D_{5/2,3/2}$ and (4,5)$F_{7/2,5/2}$ states of Mg$^+$, obtained 
by using our recommended data and the experimental energies, are listed in Table~\ref{tabMg}.  
Our estimated lifetime values for $3P_{3/2,1/2}$ states and most of the excited states are in perfect agreement with theoretical MCHF calculations by Fischer et al.~\cite{fischer} and weakest bound electron potential model theory (WBEPMT) calculations by Celik et al.~\cite{celik} except for the discrepancy that our results for $5P_{3/2,1/2}$ states do not match very well with the values given by Fischer et al.~\cite{fischer}. Discrepancy is also found between our lifetime  result of $5D_{3/2}$ state and other theoretical calculations ~\cite{celik,fischer}. We have employed more sophisticated relativistic all order method whereas, MCHF and WBEPMT methods employed by  ~\cite{fischer} and ~\cite{celik} respectively are non-relativistic methods. Therefore we advocate the reliability of our results in this case.  No experimental results are available for the lifetimes of these states and we propose  experimental analysis in future.  For the $3P_{3/2,1/2}$ states, a perfect agreement can be noticed between our calculated values with those given by Majumder et al.~\cite{majumder2002}, which are  calculated from the values of transition probabilities given therein, and with Curtis et al.~\cite{curtis1968}. It is an important point to be noted that our calculated lifetimes for the $3P_{3/2,1/2}$ states show excellent agreement with the available experimental values by Smith et al.~\cite{smith1966} and Ansbacher et.al.~\cite{ansbacher1989}. Also, our theoretical lifetime value for the  $3P_{3/2}$ state shows an excellent agreement with the experimental value reported by Herrmann et al.~\cite{herrmann2009}.

\textit{Ca$^+$ ion}: In Table \ref{tab3}, we list and compare our lifetime data with the available theoretical 
results~\cite{safronova2011,sahoo2006,guet1991,poirier, brage,kreuter,tang2013,safsaf2011,arora2007,sahoo2009,theodosiou1989} and experimental 
values~\cite{kreuter,barton,lidberg,knoop,jin1993,ansbacher1985,smith1966,emmoth1975,andersen1970,donald}. The lifetimes of the $3D_{5/2,3/2}$ states have 
contributions entirely from the E2 channel, whereas the M1 transition rates are neglected due to very small energy difference between the $3D_{5/2}$ 
and $3D_{3/2}$ states. From Table \ref{tab3}, one can notice remarkable agreement between our theoretical results for the $3D_{5/2}$ and $3D_{3/2}$ 
states' lifetimes with other theoretical calculations by Safronova et al. \cite{safronova2011}, who have also used the all-order method, Guet et 
al. \cite{guet1991} who have used RMBPT, Poirier et al. \cite{poirier} who have used Semi-empirical model and Brage et al. \cite{brage} who have 
used MCHF method. However, discrepancies are noticed from the RCC calculations by Sahoo et al.~\cite{sahoo2006}, all-order calculations by Kreuter
et al.~\cite{kreuter} and relativistic structure calculations modeled with semi-empirical fixed core potential by Tang et al.~\cite{tang2013}. 
Disagreement between our calculations  and  all-order calculations by Kreuter et al.~\cite{kreuter} is attributed to the fact that in the 
present work the values from the SD method are used for the calculation of the lifetimes of these states whereas in Ref.~\cite{kreuter} scaled values
of the SD matrix element were used. Our theoretical lifetime value  for $3D_{3/2}$ state is in very close agreement with the experimental value given by Lidberg et al.~\cite{lidberg}. However, we notice small differences between our theoretical values and the other experimental results 
for the lifetimes of the $3D_{5/2}$ and $3D_{3/2}$ states. The experimental results from various measurements for the lifetimes of the $3D_{5/2}$ and $3D_{3/2}$ states are 
themselves not in very good agreement with each other; thus, it does not allow us to make a firm conclusion if our theoretical data is in better 
agreement with the experimental values for these states.  Also, our theoretical calculations for the lifetimes of the $5P_{3/2,1/2}$ states are in 
very good agreement with the other theoretical calculations. Our results for $4P_{3/2,1/2}$ states match very well with all the theoretical results available and the measurements by Ansbacher 
et.al~\cite{ansbacher1985}, which were performed using the Pulse laser excitation method and with the measurement by Smith et.al. \cite{smith1966} 
using the Hanle effect. Recent lifetime measurements have been reported by Hettrich et al.~\cite{hettrich2015}  and Meir et al.~\cite{meir2020}  for the 4$P_{1/2}$ and 4$P_{3/2}$ states respectively. Our results are in accordance with these measurements within experimental error bars. However, our theoretical lifetime results for the $4P_{3/2,1/2}$ states differ from the experimental results reported by Jin et al.~\cite{jin1993} by 3\%. The theoretical results by Safronova et al. ~\cite{safsaf2011} and the more recent experimental results by Hettrich et al.~\cite{hettrich2015} and Meir et al.~\cite{meir2020} also report this discrepancy with the Jin et al.~\cite{jin1993} for the lifetimes of the $4P_{3/2,1/2}$ states which supports the fact that our values are reliable. Lifetimes of high-lying excited states for this ion have been calculated by 
Safronova et al.~\cite{safsaf2011} and Theodosiou~\cite{theodosiou1989}. The calculations by Theodosiou, who used Hartree-Slater method, over 
estimate the lifetimes of these states. However, to the best of our knowledge  not many lifetime measurements have been performed for the high-lying 
states of this ion. There are only a few experimental results for comparison such as for the excited $5S_{1/2}$, $4D_{5/2,3/2}$ and $5D_{3/2}$ states 
reported by Andersen et al.~\cite{andersen1970} by employing the beam-foil technique and for the $4D_{3/2}$ and $5S_{1/2}$ states by Emmoth et al.~\cite{emmoth1975} again using the beam-foil technique with cascading effects.   
These experimental measurements are in good agreement with our theoretical results for these states except for a small disagreement between the measurement in Ref.~\cite{emmoth1975} for the $4D_{3/2}$ state.  However, our result for this state and other higher excited states is in excellent agreement with calculations done by Safronova et al.~\cite{safsaf2011} which emphasizes the correctness of our results.

\textit{Sr$^+$ ion}: We list the lifetimes of the (4-7)$D_{5/2,3/2}$, (5-8)$P_{3/2,1/2}$, (6-8)$S_{1/2}$ and (4,5)$F_{7/2,5/2}$ states of Sr$^+$ in
Table~\ref{tab3}. These lifetimes are obtained using the transitions rates listed in Table~\ref{parameterssr} except for the lifetimes of the 
metastable $4D_{5/2,3/2}$ states, which have been calculated using the transition rates due to the E2 and M1 channels given in 
Tables~\ref{parameterse2} and \ref{parametersm1} respectively. We compare these lifetimes with the available theoretical calculations by Safronova et 
al. using relativistic all-order method~\cite{safronova2011}, Sahoo et al. using RCC method~\cite{sahoo2006}, Poirier et al. using Semi-empirical 
approach~\cite{poirier}, Guet \&Johnson using RMBPT~\cite{guet1991} and Filippin et al. using semi-empirical core potential approach~\cite{filippin}. Our results are in good agreement with all other theoretical calculations. Although 
one can notice few differences in some levels. The largest disagreement of about 10\% between our results and the calculations by Sahoo et al.~\cite{sahoo2006} is observed for the $4D_{5/2}$ state calculated using the RCCSD method including nonlinear terms. We notice that there is larger disagreement of our 4$D_{j}$ state lifetimes with the experiments done by Gerz et al.~\cite{gerz} in 1987 but good agreement of our theoretical values with more recent experimental measurements by Biemont et al.~\cite{biemont}, Mannervik et al.~\cite{mannervik} and
Madej et al.~\cite{madej} can be noticed for these states which indicates the reliability of our results. Also, our calculated values for the $5P_{j}$ and $4F_{j}$ states are in excellent accord with the measurements by Gallagher et al.~\cite{gallagher1967}, Pinnington et al.~\cite{pinnington1995} and Kuske et al.~\cite{kuske1978}. The  calculations for the lifetimes of other higher levels in Sr$^+$ will be helpful in further assessing the accuracy of our calculated 
results.

\textit{Ba$^+$ ion}: We carried out calculations for the lifetimes of 12 excited levels in Ba$^+$ which are presented in Table~\ref{tab5} along with 
comparison of our results with other theoretical and experimental values. Our value for the $5D_{3/2}$ lifetime within uncertainty limits agree excellently with other theoretical calculations given in Refs.~\cite{safronova2011, guet1991, gurell2007, Dzuba2001} except a small discrepancy with theoretical values given in Refs. ~\cite{sahoo2006, gopakumar}. Similarly, for the $5D_{5/2}$ and $6P_{1/2}$ states, our value is in very good agreement with all other theoretical lifetimes except for slight disagreement with Refs.~\cite{guet1991, gopakumar} and for $6P_{3/2}$ state the only disagreement is with Ref.~\cite{guet1991}. Our lifetime calculations for the $5D_{5/2,3/2}$ and $6P_{3/2,1/2}$ states agree perfectly with available experimental values within the experimental uncertainties. For the $5D_{5/2,3/2}$ state additional measurements have been performed by Gurell et al.~\cite{gurell2007}, Auchter et al.~\cite{auchter2014} and Zhang et al.~\cite{zhang2020}.
Our calculations are well within the uncertainty limits of these experimental values. Our theoretical results for the $6P_{3/2,1/2}$ states match well with two recent measurements by  Arnold et al. ~\cite{arnold2019} and Zhang et al. ~\cite{zhang2020} within experimental uncertainties. For the 
$7P_{3/2,1/2}$ states, discrepancies can be noticed between our calculations and the experimental values given in Ref.~\cite{pinnington1995}. There remains a scarcity of theoretical and experimental lifetime data for many of the higher excited states. In order to further test the accuracy of the lifetimes of the $7P_{3/2,1/2}$ states, we would like to suggest to carry out more experiments as there are not enough reliable measurements of 
the lifetimes of these states available in the literature.

\begin{longtable}{@{\extracolsep\fill}rlllllllllllll@{}}
\caption{\label{matrixmg} Reduced E1 matrix elements (in a.u.) of the Mg$^+$ alkaline earth metal ion calculated by using the DHF, SD and SDpT methods along with SD and SDpT velocity gauge matrix elements. The additional label ``sc'' indicates the scaled values. Last column lists the ${\rm{\textit{O}}}^{\rm{{Final}}}$ recommended values of matrix elements. The absolute uncertainties (${\rm{\delta}}$) are given in the last column of the table.}\\
\hline
\hline
Transition & ${\rm{\textit{O}}}^{\rm{DHF}}$ & \multicolumn{2}{c}{${\rm{\textit{O}}}^{\rm{SD}}$} &  ${\rm{\textit{O}}}^{\rm{SD}}_{\rm{sc}}$ & \multicolumn{2}{c}{${\rm{\textit{O}}}^{\rm{SDpT}}$}& ${\rm{\textit{O}}}^{\rm{SDpT}}_{\rm{sc}}$ & ${\rm{\textit{O}}}^{\rm{Final}}$ & \textbf{${\rm{\delta}}$}\\
\cline{3-4}\cline{6-7}
&  & L Gauge & V Gauge & & L Gauge & V Gauge & & \\
\hline
\hline
$  3P_{1/2}-3S_{1/2}          $ &2.462&  2.369 &2.369 &  2.369  &  2.369&2.369 & 2.419 & 2.37 & 0.05 \\
$  3P_{3/2}-3S_{1/2}          $ &3.482&  3.351 &3.351 &  3.351  &  3.351&3.351 & 3.415 & 3.35 & 0.06\\
$  4S_{1/2}-3P_{1/2}          $ &1.706&  1.693 &1.692 &  1.694  &  1.693&1.692 & 1.694 &  1.694 & 0.001\\
$  4S_{1/2}-3P_{3/2}          $ &2.421&  2.404 &2.402 &  2.404  &  2.404&2.402 & 2.365 & 2.40 & 0.03\\
$  3D_{5/2}-3P_{3/2}          $ &5.734&  5.587 &5.587 &  5.587  &  5.587&5.586 & 5.629 & 5.58   &0.04\\
\hline
\hline
\end{longtable}

\begin{longtable}{@{\extracolsep\fill}rlllllllllll@{}}
\caption{\label{matrixca} Reduced E1 matrix elements (in a.u.) of the Ca$^+$ alkaline earth metal ion calculated by using the DHF, SD and SDpT methods along with SD and SDpT velocity gauge matrix elements and the label ``sc'' indicates the scaled values. Last column lists the ${\rm{\textit{O}}}^{\rm{{Final}}}$ recommended values of matrix elements. The absolute uncertainties (${\rm{\delta}}$) are given in the last column of the table.}\\
\hline
\hline
Transition   & ${\rm{\textit{O}}}^{\rm{DHF}}$ &  \multicolumn{2}{c}{${\rm{\textit{O}}}^{\rm{SD}}$} &  ${\rm{\textit{O}}}^{\rm{SD}}_{\rm{sc}}$ & \multicolumn{2}{c}{${\rm{\textit{O}}}^{\rm{SDpT}}$} & ${\rm{\textit{O}}}^{\rm{SDpT}}_{\rm{sc}}$ & ${\rm{\textit{O}}}^{\rm{Final}}$ & \textbf{${\rm{\delta}}$}\\
\cline{3-4}\cline{6-7}
&  & L Gauge & V Gauge & & L Gauge & V Gauge & & \\
\hline
\hline
$  4P_{1/2}-4S_{1/2}$ &3.201&  2.898& 2.897 &  2.907 &  2.913& 2.914 &2.907 & 2.89 &0.02\\
$  4P_{1/2}-3D_{3/2}$ &3.082&  2.417& 2.270 &  2.464 &  2.468& 2.288 &2.450 & 2.46 &0.01\\
$  4P_{3/2}-4S_{1/2}$ &4.526&  4.099& 4.098 &  4.112 &  4.120& 4.122 &4.112 & 4.09 &0.02\\
$  4P_{3/2}-3D_{3/2}$ &1.376&  1.079&1.014  &  1.100 &  1.101&1.022  &1.094 & 1.100 &0.006\\
$  4P_{3/2}-3D_{5/2}$ &4.134&  3.245& 3.048 &  3.306 &  3.313& 3.073 &3.288 & 3.30 &0.02\\
\hline
\hline
\end{longtable}

\begin{longtable}{@{\extracolsep\fill}rlllllllllll@{}}
\caption{\label{matrixsr} Reduced E1 matrix elements (in a.u.) of the Sr$^+$ alkaline earth metal ion calculated by using the DHF, SD and SDpT 
methods along with SD and SDpT velocity gauge matrix elements. Values with the label ``sc'' indicate the scaled values. Last column lists the ${\rm{O}}^{\rm{{Final}}}$ recommended values of matrix 
elements. The absolute uncertainties (${\rm{\delta}}$) are given in the last column of the table.}\\
\hline
\hline
Transition   & ${\rm{\textit{O}}}^{\rm{DHF}}$ &  \multicolumn{2}{c}{${\rm{\textit{O}}}^{\rm{SD}}$} &  ${\rm{\textit{O}}}^{\rm{SD}}_{\rm{sc}}$ & \multicolumn{2}{c}{${\rm{\textit{O}}}^{\rm{SDpT}}$} & ${\rm{\textit{O}}}^{\rm{SDpT}}_{\rm{sc}}$ & ${\rm{\textit{O}}}^{\rm{Final}}$ & \textbf{${\rm{\delta}}$} \\
\cline{3-4}\cline{6-7}
&  & L Gauge & V Gauge & & L Gauge & V Gauge & & \\
\hline
\hline
$  5P_{1/2}-5S_{1/2}$ &3.484&  3.078 &3.079 &  3.091   &  3.097&3.099  &3.090 & 3.07 &0.02\\
$  5P_{1/2}-4D_{3/2}$ &3.729&  3.083 & 2.893&  3.112   &  3.119& 2.898 &3.102 & 3.11 &0.03\\
$  5P_{3/2}-5S_{1/2}$ &4.921&  4.351 &4.351 &  4.368   &  4.377& 4.379 &4.367 & 4.35 &0.03\\
$  5P_{3/2}-4D_{3/2}$ &1.657&  1.369 &1.289 & 1.383    &  1.386& 1.293 &1.378 & 1.38 &0.01\\
$  5P_{3/2}-4D_{5/2}$ &5.002&  4.150 &3.901 & 4.187    &  4.198& 3.909 &4.173 & 4.18 &0.04\\
\hline
\hline
\end{longtable}

\begin{longtable}{@{\extracolsep\fill}rlllllllllll@{}}
\caption{\label{matrixba} Reduced E1 matrix elements (in a.u.) of the Ba$^+$ alkaline earth metal ion calculated by using the DHF, SD and SDpT 
methods along with SD and SDpT velocity gauge matrix elements. Values with the label ``sc'' indicate the scaled values. Last column lists the ${\rm{\textit{O}}}^{\rm{{Final}}}$ recommended values of matrix elements. The absolute uncertainties (${\rm{\delta}}$) are given in the last column of the table.}\\
\hline
\hline
Transition  &${\rm{\textit{O}}}^{\rm{DHF}}$ & \multicolumn{2}{c}{${\rm{\textit{O}}}^{\rm{SD}}$} &  ${\rm{\textit{O}}}^{\rm{SD}}_{\rm{sc}}$&\multicolumn{2}{c}{${\rm{\textit{O}}}^{\rm{SDpT}}$} & ${\rm{\textit{O}}}^{\rm{SDpT}}_{\rm{sc}}$ &${\rm{\textit{O}}}^{\rm{Final}}$ & \textbf{${\rm{\delta}}$} \\
\cline{3-4}\cline{6-7}
&  & L Gauge & V Gauge & & L Gauge & V Gauge & & \\
\hline
\hline
$  6P_{1/2}-6S_{1/2}$ &3.890 & 3.338& 3.341 &  3.358    &  3.371& 3.374 & 3.358 & 3.33 & 0.03\\
$  6P_{1/2}-5D_{3/2}$ &3.745 & 3.055& 2.909  &  3.077    &  3.100& 2.910 & 3.066 & 3.07 & 0.02 \\
$  6P_{3/2}-6S_{1/2}$ &5.477 & 4.710& 4.714 &  4.739    &  4.757&4.758 & 4.738 & 4.71 & 0.05\\
$  6P_{3/2}-5D_{3/2}$ &1.635 & 1.334& 1.270 &  1.344    &  1.355& 1.274 & 1.339 & 1.34 & 0.01\\
$  6P_{3/2}-5D_{5/2}$ &5.001 & 4.111& 3.909 &  4.139    &  4.171& 3.915 & 4.124 & 4.13 & 0.03\\
\hline
\hline
\end{longtable}

\begin{longtable}{@{\extracolsep\fill}rllllll@{}}
\caption{\label{matrixe2m1} Reduced E2 and M1 matrix elements (in a.u.) of the metastable states in the Ca$^+$, Sr$^+$ and Ba$^+$ alkaline earth metal
ion calculated by using the DHF and SD methods labeled as ${\rm{\textit{O}}}^{\rm{DHF}}$ and ${\rm{\textit{O}}}^{\rm{Final}}$, respectively. The absolute uncertainties (${\rm{\delta}}$) are also given in the table. }\\
Transition & \multicolumn{3}{c}{E2 Amplitudes} & \multicolumn{3}{c}{M1 Amplitudes} \\
\cline{2-4}\cline{5-6}
& {${\rm{\textit{O}}}^{\rm{DHF}}$} & {${\rm{\textit{O}}}^{\rm{Final}}$} & \textbf{${\rm{\delta}}$} & {${\rm{\textit{O}}}^{\rm{DHF}}$} & {${\rm{\textit{O}}}^{\rm{Final}}$} \\
\hline
\hline
Ca$^+$\\
 $3D_{3/2}-4S_{1/2}$ &  9.767 &  7.78&0.04& 0.0050& 0.0000\\
 $3D_{5/2}-4S_{1/2}$ & 11.978 &  9.56&0.04& - &-\\
 $3D_{5/2}-3D_{3/2}$ &  5.018 &  3.68&0.02& 1.5421&1.5491\\

Sr$^+$\\
 $4D_{3/2}-5S_{1/2}$ & 12.968& 11.01&0.04& 0.0093 &0.0000\\
 $4D_{5/2}-5S_{1/2}$ & 15.972& 13.60&0.05& -&-\\
 $4D_{5/2}-4D_{3/2}$ &  3.565&  5.90&0.02& 1.5385&1.5492\\

Ba$^+$\\
 $5D_{3/2}-6S_{1/2}$ &  8.510& 12.49&0.11 &0.0140&0.0002\\
 $5D_{5/2}-6S_{1/2}$ & 10.650& 15.65&0.14 &-&-\\
 $5D_{5/2}-5D_{3/2}$ &  4.814&  6.65&0.06 &1.5315&1.5493\\
 
\hline\\		      
\end{longtable}

\vspace{0.1 cm}
\begin{longtable}{@{\extracolsep\fill}rllllllll@{}}
\caption{\label{parametersmg} Wavelengths $\lambda$ (in {\AA}), transition probabilities $A_{vk}$ (in $\rm{s^{-1}}$), absorption oscillator strengths $f_{kv}$ 
and line strengths $S_{vk}$ (in a.u.) of the Mg$^+$ alkaline earth metal ion due to E1 channel are presented with their absolute uncertainties. Values given in square brackets represent the power 
of 10. $A_{vk}$ and $f_{kv}$ values are compared with results given in NIST ASD~\cite{kramida}.}\\
\hline
\hline
\multicolumn{2}{c}{\textbf{State}} & {\textbf{$\lambda$}}  & \textbf{$S_{vk}$} & \multicolumn{2}{c}{\textbf{$A_{vk}$}} & \multicolumn{2}{c}{\textbf{$f_{kv}$}} &  \\ 
\cline{1-2}\cline{5-6}\cline{7-8}
Upper $(v)$ & Lower $(k)$ && &Present & Ref.~\cite{kramida} & Present &Ref.~\cite{kramida} &\\
\hline
\hline
$ 3P_{1/2} $&$ 3S_{1/2} $ & 2803.531 & 5.61$\pm$0.24[0]   & 2.58$\pm$0.11[8] & 2.57[8]& 3.04$\pm$0.13[-1]& 3.03[-1]&\\
$ 3P_{3/2} $&$ 3S_{1/2} $ & 2796.351 & 1.12$\pm$0.04[1]   & 2.60$\pm$0.09[8] & 2.60[8]& 6.09$\pm$0.19[-1]& 6.08[-1]&\\
$ 4S_{1/2} $&$ 3P_{1/2} $ & 2929.489 & 2.869$\pm$0.003[0] & 1.156$\pm$0.001[8] & 1.15[8]& 1.487$\pm$0.002[-1]& 1.48[-1]&\\
$ 4S_{1/2} $&$ 3P_{3/2} $ & 2937.369 & 5.78$\pm$0.18[0]   & 2.31$\pm$0.09[8] & 2.30[8]& 1.49$\pm$0.05[-1]& 1.49[-1]&\\
$ 3D_{5/2} $&$ 3P_{3/2} $ & 2798.823 & 3.12$\pm$0.05[1]   & 4.81$\pm$0.07[8] & 4.79[8]& 8.46$\pm$0.08[-1]& 8.44[-1]&\\
\end{longtable}

\vspace{0.1 cm}
\begin{longtable}{@{\extracolsep\fill}rlllllllll@{}}
\caption{\label{parametersca} Properties such as wavelengths $\lambda$ (in {\AA}), transition probabilities $A_{vk}$ (in $\rm{s^{-1}}$), absorption oscillator 
strengths $f_{kv}$ and line strengths $S_{vk}$ (in a.u.) due to E1 channel in the Ca$^+$ alkaline earth metal ion are presented with their absolute uncertainties. Values given in square brackets represent the power of 10. 
$A_{vk}$ and $f_{kv}$ values are compared with results given in NIST ASD~\cite{kramida}.}\\
\hline
\hline
\multicolumn{2}{c}{\textbf{State}} & {\textbf{$\lambda$}}  & \textbf{$S_{vk}$} & \multicolumn{2}{c}{\textbf{$A_{vk}$}} & \multicolumn{2}{c}{\textbf{$f_{kv}$}} & \\ 
\cline{1-2}\cline{5-6}\cline{7-8}
Upper $(v)$ & Lower $(k)$ && &Present &  Ref.~\cite{kramida}  & Present &  Ref.~\cite{kramida}  \\
\hline
\hline
$ 4P_{1/2} $&$ 4S_{1/2} $ & 3969.591 & 8.39$\pm$0.09[0] & 1.36$\pm$0.01[8]  & 1.4[8]  &3.21$\pm$0.03[-1] & 3.3[-1]&\\
$ 4P_{1/2} $&$ 3D_{3/2} $ & 8664.520 & 6.07$\pm$0.07[0] & 0.95$\pm$0.01[7]  & 1.06[7] &5.32$\pm$0.06[-2]&5.97[-2]&\\
$ 4P_{3/2} $&$ 4S_{1/2} $ & 3934.777 & 1.68$\pm$0.02[1] & 1.39$\pm$0.01[8]  & 1.47[8] &6.48$\pm$0.07[-1]& 6.82[-1]&\\
$ 4P_{3/2} $&$ 3D_{3/2} $ & 8500.358 & 1.21$\pm$0.01[0] & 0.99$\pm$0.01[6]  & 1.11[6] &1.08$\pm$0.01[-2]& 1.20[-2]&\\
$ 4P_{3/2} $&$ 3D_{5/2} $ & 8544.437 & 1.09$\pm$0.01[1] & 8.87$\pm$0.09[6]  & 9.9[6]  &6.47$\pm$0.07[-2]& 7.2[-2]&\\
\end{longtable}

\vspace{1 cm}
\begin{longtable}{@{\extracolsep\fill}rllllllll@{}}
\caption{\label{parameterssr} Wavelengths $\lambda$ (in {\AA}), transition probabilities $A_{vk}$ (in $\rm{s^{-1}}$), absorption oscillator strengths $f_{kv}$ 
and line strengths $S_{vk}$ (in a.u.) in the Sr$^+$ alkaline earth metal ion due to E1 channel are presented with their absolute uncertainties. Values given in square brackets represent the power of 10. $A_{vk}$ and $f_{kv}$ values are compared with results given in NIST ASD~\cite{kramida}.}\\
\hline
\hline
\multicolumn{2}{c}{\textbf{State}} & {\textbf{$\lambda$}}  & \textbf{$S_{vk}$} & \multicolumn{2}{c}{\textbf{$A_{vk}$}} & \multicolumn{2}{c}{\textbf{$f_{kv}$}} &  \\ 
\cline{1-2}\cline{5-6}\cline{7-8}
Upper $(v)$ & Lower $(k)$ && &Present & Ref.~\cite{kramida} & Present & Ref.~\cite{kramida} &\\
\hline
\hline
$ 5P_{1/2} $&$ 5S_{1/2} $ & 4216.706  & 9.47$\pm$0.11[0]  &1.28$\pm$0.02[8]& 1.28[8]&3.41$\pm$0.04[-1]& 3.41[-1]&\\
$ 5P_{1/2} $&$ 4D_{3/2} $ & 10917.876 & 9.68$\pm$0.18[0]  &7.53$\pm$0.14[6]& 7.46[6]&6.73$\pm$0.12[-2]& 6.67[-2]&\\ 
$ 5P_{3/2} $&$ 5S_{1/2} $ & 4078.861  & 1.89$\pm$0.02[1]  &1.41$\pm$0.01[8]& 1.41[8]&7.05$\pm$0.08[-1]& 7.03[-1]&\\
$ 5P_{3/2} $&$ 4D_{3/2} $ & 10039.404 & 1.91$\pm$0.38[0]  & 0.96$\pm$0.02[6]& 1.0[6]&1.45$\pm$0.03[-2]& 1.5[-2]&\\ 
$ 5P_{3/2} $&$ 4D_{5/2} $ & 10330.141 & 1.75$\pm$0.03[1]  & 8.1$\pm$0.1[6]& 8.7[6]&8.6$\pm$0.1[-2]& 9.3[-2]&\\
\end{longtable}

\vspace{1 cm}
\begin{longtable}{@{\extracolsep\fill}rllllllll@{}}
\caption{\label{parametersba} Wavelengths $\lambda$ (in {\AA}), transition probabilities $A_{vk}$ (in $\rm{s^{-1}}$), absorption oscillator strengths $f_{kv}$ 
and line strengths $S_{vk}$ (in a.u.) in the Ba$^+$ alkaline earth metal ion due to the E1 channel are presented with their absolute uncertainties. Values given in square brackets represent the power of 10. $A_{vk}$ and $f_{kv}$ values are compared with results given in NIST ASD~\cite{kramida}.}\\
\hline
\hline
\multicolumn{2}{c}{\textbf{State}} & {\textbf{$\lambda$}}  & \textbf{$S_{vk}$} & \multicolumn{2}{c}{\textbf{$A_{vk}$}} & \multicolumn{2}{c}{\textbf{$f_{kv}$}} &  \\ 
\cline{1-2}\cline{5-6}\cline{7-8}
Upper $(v)$ & Lower $(k)$ && &Present & Ref.~\cite{kramida} & Present & Ref.~\cite{kramida} \\
\hline
\hline
$ 6P_{1/2} $&$ 6S_{1/2} $ & 4935.453 & 1.11$\pm$0.02[1] & 9.38$\pm$0.18[7] & 9.53[7]& 3.43$\pm$0.06[-1]& 3.48[-1]&\\ 
$ 6P_{1/2} $&$ 5D_{3/2} $ & 6498.693 & 9.46$\pm$0.14[0] & 3.49$\pm$0.05[7] & 3.10[7]& 11.06$\pm$0.16[-2]& 9.81[-2]&\\  
$ 6P_{3/2} $&$ 6S_{1/2} $ & 4555.309 & 2.22$\pm$0.04[1] & 1.18$\pm$0.02[8] & 1.11[8]& 7.39$\pm$0.14[-1]& 6.9[-1]&\\
$ 6P_{3/2} $&$ 5D_{3/2} $ & 5855.297 & 1.81$\pm$0.02[0] & 4.55$\pm$0.07[6] & 6.00[6]& 2.34$\pm$0.03[-2]& 3.08[-2]&\\ 
$ 6P_{3/2} $&$ 5D_{5/2} $ & 6143.413 & 1.71$\pm$0.02[1] & 3.74$\pm$0.05[7] & 4.12[7]& 1.412$\pm$0.002[-1]& 1.55[-1]&\\
\end{longtable}

\begin{longtable}{@{\extracolsep\fill}rllllllll@{}}
\caption{Comparison of some of our determined E1 transition probabilities  ($10^{8}$ per sec) in Mg$^+$ with other available theoretical and experimental data.}\\
\hline
\multicolumn{2}{c}{{State}} & {{Present}}  & {{Theory(others)}} & {{Experiment}} \\ 
\cline{1-2}
Upper $(v)$ & Lower $(k)$ &  \\
\hline
\hline
$3P_{1/2}$ & $3S_{1/2}$ & 2.58$\pm$0.11 &2.576~\cite{majumder2002}, 2.546$\pm$0.042~\cite{theodosiou1999} & 2.579$\pm$0.017~\cite{ansbacher1989}\\
$3P_{3/2}$ & $3S_{1/2}$ & 2.60$\pm$0.09 &2.597~\cite{majumder2002}, 2.563$\pm$0.043~\cite{theodosiou1999} & 2.602$\pm$0.025~\cite{ansbacher1989} \\
           &            &               &    & 2.604$\pm$0.067~\cite{herrmann2009} \\
$4P_{1/2}$ & $3S_{1/2}$ & 0.014$\pm$0.001 &0.0144~\cite{majumder2002}, 0.0154$\pm$0.0002~\cite{theodosiou1999},0.015$\pm$0.001~\cite{sofia} & 0.0138$\pm$0.0008~\cite{fitz}\\
$4P_{3/2}$ & $3S_{1/2}$ & 0.0123$\pm$0.0007 &0.0129~\cite{majumder2002} 0.0137$\pm$0.0001~\cite{theodosiou1999},0.0134$\pm$0.0004~\cite{sofia} & 0.0138$\pm$0.0008~\cite{fitz} \\
\end{longtable}

\begin{longtable}{@{\extracolsep\fill}rlllllllll@{-2pt}}
\caption{Comparison of some of our E1 transition probabilities ($10^{8}$ per sec) in Ca$^+$ with other available theoretical and experimental 
literature data.}\\
\hline
\multicolumn{2}{c}{{State}} & {{Present}}   & {{Theory(others)}}  &  {Experiment}\\ 
\cline{1-2}
Upper $(v)$ & Lower $(k)$ &  \\
\hline
\hline
$5S_{1/2}$&$4P_{3/2}$ & 1.71$\pm$0.01 &1.705\cite{safsaf2011}, 1.65~\cite{wiese}, 1.701~\cite{kurucz}, 1.687~\cite{melendez}  &  1.57$\pm$0.11~\cite{augilera}, 1.5$\pm$0.2~\cite{andersen1970}\\
$5S_{1/2}$&$4P_{1/2}$ & 0.855$\pm$0.006 &0.855\cite{safsaf2011},  0.84~\cite{wiese}, 0.867~\cite{kurucz},0.845~\cite{melendez} &  0.837$\pm$0.059~\cite{augilera}, 0.8$\pm$0.1~\cite{andersen1970}  \\
$6S_{1/2}$&$4P_{3/2}$ & 0.641$\pm$0.003 &0.642\cite{safsaf2011},  0.62~\cite{wiese}, 0.6146~\cite{kurucz}, 0.644~\cite{melendez} &  0.396$\pm$0.048~\cite{augilera}             \\
$6S_{1/2}$&$4P_{1/2}$ & 0.323$\pm$0.001 &0.323\cite{safsaf2011},  0.313~\cite{wiese}, 0.3111~\cite{kurucz}, 0.323~\cite{melendez} &  0.208$\pm$0.025~\cite{augilera}   \\
$4D_{3/2}$&$4P_{1/2}$ & 2.94$\pm$0.03 &2.944\cite{safsaf2011},  3.05~\cite{wiese}, 2.984~\cite{kurucz}, 2.85~\cite{melendez}    &  2.83$\pm$0.20~\cite{augilera},  3.0$\pm$0.2~\cite{andersen1970}\\
\end{longtable}

\begin{longtable}{@{\extracolsep\fill}rllllllll@{}}
\caption{Comparison of some of our  E1 transition probabilities ($10^{8}$ per sec) in Sr$^+$ with the available theoretical and experimental data.}\\
\hline
\multicolumn{2}{c}{{State}} & {{Present}}  & {{Theory(others)}} & {{Experiment}} \\ 
\cline{1-2}
Upper $(v)$ & Lower $(k)$ &  \\
\hline
\hline
$5P_{1/2}$&$4D_{3/2}$ & 0.075$\pm$0.001 &0.0725~\cite{guet1991}, 0.0901~\cite{brage1998}, 0.154~\cite{ziltis}  & 0.0746$\pm$0.0014~\cite{likforman2016}, 0.095$\pm$0.02~\cite{gallagher1967}\\
$5P_{3/2}$&$4D_{3/2}$ & 0.0096$\pm$0.0002 &0.0092~\cite{guet1991}, 0.0116~\cite{brage1998}, 0.018~\cite{ziltis} &0.010$\pm$0.02~\cite{gallagher1967}  \\
$5P_{3/2}$&$4D_{5/2}$ & 0.081$\pm$0.001 &0.0776~\cite{guet1991}, 0.0957~\cite{brage1998} & 0.08010$\pm$0.00089~\cite{zhang2016}, 0.087$\pm$0.015~\cite{gallagher1967} \\
$5P_{1/2}$&$5S_{1/2}$ & 1.28$\pm$0.02 &1.265~\cite{guet1991}, 1.646~\cite{brage1998}, 1.489~\cite{ziltis}   & 1.279$\pm$0.013~\cite{likforman2016}, 1.27$\pm$0.05~\cite{gallagher1967} \\
$5P_{3/2}$&$5S_{1/2}$ & 1.41$\pm$0.01 &1.396~\cite{guet1991}, 1.819~\cite{brage1998}, 1.637~\cite{ziltis}   & 1.425$\pm$0.015~\cite{zhang2016}, 1.43$\pm$0.06~\cite{gallagher1967} \\
$4F_{5/2}$&$4D_{3/2}$ & 2.87$\pm$0.07 &3.428~\cite{brage1998}, 4.583~\cite{ziltis}          & \\
$4F_{5/2}$&$4D_{5/2}$ & 0.206$\pm$0.005 &0.241~\cite{brage1998}          & \\
$4F_{7/2}$&$4D_{5/2}$ & 3.10$\pm$0.07 &3.617~\cite{brage1998}          & \\
\hline
\hline
\label{comparsr}
\end{longtable}

\begin{longtable}{@{\extracolsep\fill}rllllllll@{}}
\caption{Comparison of some of our  E1 transition probabilities  ($10^{8}$ per sec) in Ba$^+$ with other theoretical and experimental data.}\\
\hline
\multicolumn{2}{c}{{State}} & {{Present}}  & {{Theory(others)}} & {{Experiment}} \\ 
\cline{1-2}
Upper $(v)$ & Lower $(k)$ &  \\
\hline
\hline
$6P_{1/2}$&$5D_{3/2}$ & 0.349$\pm$0.005 & 0.332~\cite{safronova2011}, 0.32609~\cite{gopakumar} &  0.33$\pm$0.04~\cite{gallagher1967},  0.343$\pm$0.015~\cite{kastberg}, 0.341$\pm$0.001~\cite{arnold2019} \\
$6P_{3/2}$&$5D_{3/2}$ & 0.0455$\pm$0.0007 & 0.043~\cite{safronova2011}, 0.04255~\cite{gopakumar} &0.048$\pm$0.006~\cite{gallagher1967},  0.0469$\pm$0.0029~\cite{kastberg} \\
&  &   &    & 0.0447$\pm$0.0001~\cite{zhang2020}, 0.0445$\pm$0.0001~\cite{arnold2019}  \\
$6P_{3/2}$&$5D_{5/2}$ & 0.374$\pm$0.005 & 0.358~\cite{safronova2011}, 0.3493~\cite{gopakumar} & 0.37$\pm$0.04~\cite{gallagher1967}, 0.377$\pm$0.017~\cite{kastberg} \\
&  &   &    &  0.3677$\pm$0.0004~\cite{zhang2020}, 0.365$\pm$0.001~\cite{arnold2019}    \\
$6P_{1/2}$&$6S_{1/2}$ & 0.94$\pm$0.02 & 0.912~\cite{safronova2011}, 0.9368~\cite{gopakumar} & 0.9316$\pm$0.011~\cite{woods2010}, 0.95$\pm$0.07~\cite{gallagher1967},  0.95$\pm$0.05~\cite{kastberg}\\
$6P_{3/2}$&$6S_{1/2}$ & 1.18$\pm$0.02 & 1.154~\cite{safronova2011}, 1.1937~\cite{gopakumar}  & {1.1846$\pm$0.0014~\cite{woods2010}} 1.18$\pm$0.08~\cite{gallagher1967}, 1.06$\pm$0.04~\cite{kastberg} \\
\hline
\hline
\label{comparba}
\end{longtable}


\vspace{1 cm}

\begin{longtable}{@{\extracolsep\fill}rlllllllll@{}}
\caption{\label{parameterse2} Wavelengths $\lambda$ (in {\AA}), line strengths $S_{vk}$ (in a.u.), transition probabilities $A_{vk}$ 
(in $\rm{s^{-1}}$) and absorption oscillator strengths $f_{kv}$ in the Ca$^+$, Sr$^+$ and Ba$^+$ alkaline earth metal ions due to E2 channel. Values given in parentheses represent the power of 10.}\\
\hline
\hline
Transition & {{$\lambda$}}  & {$S_{vk}$} & \multicolumn{2}{c}{{$A_{vk}$}} &  \multicolumn{2}{c}{{$f_{kv}$}} \\ 
\cline{4-5}\cline{6-7}
 &&&Present& Others(Th, Exp) & Present& Others(Th, Exp) \\

\hline
\hline
Ca$^+$\\
$3D_{5/2}\rightarrow4S_{1/2}$   &    7293.477   & 9.141(1) & 0.827$\pm$0.008(0)& 0.8567$\pm$0.0108(0)~\cite{shao2017}    & 1.98$\pm$0.02(-8)&  2.049$\pm${0.025}(-8)~\cite{shao2017}  \\
              & & & & 0.853$\pm$0.002(0)~\cite{guan2015}  &  & 2.041$\pm${0.005}(-8)~\cite{guan2015} \\
              & & & & 0.8892$\pm$0.0049(0)~\cite{filippin} &  & 2.127$\pm${0.012}(-8)~\cite{filippin} \\
              & & & & 1.3(0)~\cite{kramida} & & 3.110(-8)~\cite{kramida} \\
\hline
\hline

\end{longtable}

\vspace{1 cm}

\begin{longtable}{@{\extracolsep\fill}rlllllllll@{}}
\caption{\label{parametersm1} Wavelengths $\lambda$ (in {\AA}), line strengths $S_{vk}$ (in a.u.), transition probabilities $A_{vk}$ 
(in $\rm{s^{-1}}$) and absorption oscillator strengths $f_{kv}$ in the Ca$^+$, Sr$^+$ and Ba$^+$ alkaline earth metal ions due to M1 channel. Values given in parentheses represent the power of 10.}\\
\hline
\hline
\multicolumn{2}{c}{{State}} & {{$\lambda$}}  & {$S_{vk}$} & \multicolumn{2}{c}{{$A_{vk}$}} &  \multicolumn{2}{c}{$f_{kv}$}  \\ 
\cline{1-2}\cline{5-6}\cline{7-8}
Upper $(v)$ & Lower $(k)$ & & & Present & Others  & Present & Others \\

\hline
\hline
Ca$^+$\\
$3D_{3/2} $ & $ 4S_{1/2}$& 7325.905 & $\sim$0& $\sim$0& 1.947$\pm$0.061(-11)~\cite{filippin} & $\sim$0&$\sim$0~\cite{filippin}  \\
$3D_{5/2} $ & $ 3D_{3/2}$& 1647717.911 & 2.399(0)& 2.412(-6)& 2.422$\pm$0.011(-6)~\cite{filippin} & 1.472(-9) & 1.478$\pm${0.006}(-9)~\cite{filippin} & \\
            &             & & &         &  2.5(-6)~\cite{guan2015}, 2.41(-6)~\cite{kramida} &   &  1.526(-9)~\cite{guan2015} \\  
            &             & & &         &                                                   &   &  1.471(-9)~\cite{kramida} \\
\hline
\hline

\end{longtable}


\begin{longtable}{@{\extracolsep\fill}rllll@{}}
\caption{Lifetimes (in s) of a few excited states of the Mg$^+$ ion from the present work and from other available literature data. The numbers in
parentheses represent powers of 10.}\\
\multicolumn{1}{c}{State} &
\multicolumn{1}{c}{Present} &
\multicolumn{1}{c}{Th (Others)} &
\multicolumn{1}{c}{Experiment} \\ 

\hline\\
$3P_{1/2}$ & 3.875$\pm$0.002(-9)& 3.881(-9)~\cite{majumder2002}, 3.872(-9)~\cite{curtis1968}, 3.85(-9)~\cite{celik}, 3.64(-9)~\cite{fischer}  &3.854$\pm$0.030(-9)~\cite{ansbacher1989} \\
$3P_{3/2}$ & 3.84$\pm$0.14(-9)&3.850(-9)~\cite{majumder2002}, 3.842(-9)~\cite{curtis1968}, 3.81(-9)~\cite{celik}, 3.61(-9)~\cite{fischer}  & {3.84$\pm$0.10(-9)~\cite{herrmann2009}}, 3.810$\pm$0.040(-9)~\cite{ansbacher1989}  \\
           &                  &     & 3.67$\pm$0.18(-9)~\cite{smith1966}  \\                                                            
$4S_{1/2}$ & 2.88$\pm$0.08(-9)& 2.90(-9)~\cite{celik}, 2.95(-9)~\cite{fischer} &  \\
$3D_{3/2}$ & 2.071$\pm$0.004(-9)& 2.04(-9)~\cite{celik}, 1.96(-9)~\cite{fischer} &  \\
\end{longtable}

\begin{longtable}{@{\extracolsep\fill}rllll@{}}
\caption{Lifetimes (in s) of a few excited states of Ca$^+$ from the present work and from other available literature data. The numbers in
parentheses represent powers of 10.}\\
\multicolumn{1}{c}{State} &
\multicolumn{1}{c}{Present} &
\multicolumn{1}{c}{Theory(others)} &
\multicolumn{1}{c}{Experiment} \\ 

\hline\\
$3D_{3/2}$ &1.24$\pm$0.01 & 1.243\cite{safronova2011},  1.271\cite{guet1991}, 1.200\cite{poirier}, 1.20\cite{brage}  & 1.20{$\pm0.01$}\cite{barton}, 1.17$\pm0.05$\cite{lidberg} \\
           &      &1.185$\pm0.007$\cite{sahoo2006}, 1.143$\pm0.001$\cite{tang2013}, 1.196$\pm0.011$\cite{kreuter}    &  1.176$\pm0.011$\cite{kreuter} \\
$3D_{5/2}$ &1.21$\pm$0.01 & 1.209\cite{safronova2011}, 1.236\cite{guet1991}, 1.165$\pm0.011$(-9)\cite{kreuter} & 1.168$\pm0.009$\cite{kreuter}, 1.168$\pm0.007$~\cite{barton}\\ 
           &      &  1.110$\pm0.009$\cite{sahoo2006}, 1.170\cite{poirier}, 1.114$\pm0.001$\cite{tang2013}, 1.163~\cite{brage} &  1.152$\pm0.020$~\cite{knoop}, 1.177$\pm0.010$\cite{donald}\\
$4P_{1/2}$ & 6.87$\pm$0.06(-9) & 6.88{$\pm0.06$}(-9)\cite{safsaf2011}, 6.870(-9)\cite{poirier}, 6.875(-9)\cite{arora2007},  & {6.904$\pm0.026$~\cite{hettrich2015}}, 6.96$\pm0.35$(-9)\cite{ansbacher1985}\\
\end{longtable}

\begin{longtable}{@{\extracolsep\fill}rlll@{}}
\caption{Lifetimes (in s) of a few excited states of Sr$^+$ from the present work and from other available literature data. The numbers in
parentheses represent powers of 10.}\\
\multicolumn{1}{c}{State} &
\multicolumn{1}{c}{Present} &
\multicolumn{1}{c}{Theory(others)} &
\multicolumn{1}{c}{Experiment} \\ 

\hline\\
$4D_{3/2}$ & 0.451$\pm$0.003  &0.445$\pm$0.003~\cite{filippin}, 0.4509\cite{safronova2011}, 0.454\cite{guet1991} & 0.435$\pm0.004$\cite{mannervik}, 0.435$\pm0.004$\cite{biemont}, 0.395$\pm0.038$\cite{gerz} \\
           &        &   0.441\cite{poirier}, 0.426$\pm0.008$\cite{sahoo2006}  &  \\
$4D_{5/2}$ & 0.403$\pm$0.003 & 0.398$\pm$0.003~\cite{filippin}, 0.4029\cite{safronova2011}, 0.405\cite{guet1991}  & 0.372$\pm0.025$\cite{madej}, 0.408$\pm$0.022~\cite{biemont}, 0.345$\pm0.033$~\cite{gerz} \\
           &        &  0.396\cite{poirier}, 0.357$\pm$0.012\cite{sahoo2006}  &  \\
$5P_{1/2}$ & 7.38$\pm$0.08(-9) &7.383(-9)\cite{safronova2011}, 7.48(-9)\cite{guet1991},7.43(-9)\cite{poirier} & 7.39$\pm$0.07(-9)\cite{pinnington1995}, 7.47$\pm$0.07(-9)\cite{kuske1978}, 7.35$\pm$0.03(-9)~\cite{gallagher1967}  \\
\end{longtable}

\begin{longtable}{@{\extracolsep\fill}rlll@{}}
\caption{Lifetimes (in s) of a few excited states of Ba$^+$ from the present work and from other available literature data. The numbers in
parentheses represent powers of 10.}\\
\multicolumn{1}{c}{State} &
\multicolumn{1}{c}{Present} &
\multicolumn{1}{c}{Th(others)} &
\multicolumn{1}{c}{Experiment} \\ 

\hline\\
$5D_{3/2}$ &83.00$\pm$1.00& 83.26\cite{safronova2011},  83.7\cite{guet1991}, 82.0~\cite{gurell2007}& 79.8$\pm4.6$\cite{yu1997}, 89.4$\pm$15.6~\cite{gurell2007}\\
           &     &   81.5\cite{Dzuba2001}, 81.4~\cite{gopakumar}, 80.086$\pm$0.714\cite{sahoo2006}   &   \\
$5D_{5/2}$ &30.81$\pm$0.65& 30.88~\cite{safronova2011}, 30.3\cite{Dzuba2001}, 31.6~\cite{gurell2007},  & {30.14$\pm$0.40~\cite{zhang2020}},  32.0$\pm$4.6~\cite{gurell2007}, 31.2$\pm$0.9~\cite{auchter2014}\\
           &     &   37.2\cite{guet1991},36.5~\cite{gopakumar}, 29.856$\pm$0.296\cite{sahoo2006}     & 34.5$\pm$3.5\cite{madejPRA}, 47$\pm$16~\cite{plumelle}  \\
$6P_{1/2}$ & 7.76$\pm$0.11(-9)  &7.79(-9)~\cite{safronova2011}, 7.99(-9)\cite{guet1991}, 7.89(-9)\cite{Dzuba2001} &  7.74$\pm0.40$(-9)\cite{gallagher1967}, 7.90$\pm$0.10(-9)\cite{pinnington1995} \\
\end{longtable}

\section{Conclusion} \label{cons}

In the present work, electric dipole matrix elements between many excited states of the alkaline earth ions are reported by analyzing results from 
four different approximations in the singles and doubles all-order theory and accounting for the valence triple excitations in the perturbative 
approach. The final values are recommended based on the correlation contributions to energies to obtain the values agreeing with the experimental 
results. This includes 84 transitions in Mg$^+$, 84 transitions in Ca$^+$, 104 transitions in Sr$^+$ and 60 transitions in Ba$^+$. By combining our
calculated electric dipole matrix elements with the experimental values, we have also estimated oscillator strengths, transition probabilities and
lifetimes of many high-lying excited states of the above ions. We have also estimated the magnetic dipole and electric quadrupole transition matrix 
elements of the metastable states of the Ca$^+$, Sr$^+$ and Ba$^+$ alkaline-earth metal ions. These results are further used to estimate the 
oscillator strengths and transition probabilities of the forbidden transitions, and the lifetimes of the metastable states. We have compared 
our results with the available theoretical and experimental literature. There were no 
data available for the comparison purpose with some of our estimated values, however, we believe that these results are of similar accuracy as the 
other results that agree well with the experimental values and calculations made using different many-body methods. The reliability of our results is additionally emphasized by the fact that the comparison of the values from the length and velocity gauge expressions of E1 matrix elements present a very good agreement with each other and differences are within the estimated uncertainties. Also, at many places our results demonstrate very good agreement with the most recent experimental measurements wherever available. All these reported results will serve as excellent benchmarks for comprehending radiative properties of the undertaken alkaline-earth metal ions. 

\ack

The work of B.A. is supported by DST-SERB(India) Grant No. EMR/2016/001228. The employed all order method was developed in the group of Professor M. S. Safronova of the University of Delaware, USA.

\clearpage

\end{document}